\documentclass{elsart}
\usepackage{graphics}
\usepackage{graphicx}
\usepackage{amssymb,amsmath}
\usepackage[latin1]{inputenc}
\newcommand{\mgb}{MgB$_2$}
\begin{document}

\begin{frontmatter}

\title{Depinning frequency in a heavily neutron-irradiated \mgb\ sample}

\author[label1]{M. Bonura},
\author[label1]{A. Agliolo Gallitto},
\author[label1,cor1]{M. Li Vigni},
\author[label2]{A. Martinelli},
\address[label1]{CNISM and Dipartimento di Scienze Fisiche e Astronomiche,
Universit\`{a} di Palermo, via Archirafi 36, 90123 Palermo, Italy}
\address[label2]{CNR-INFM-LAMIA and Dipartimento di Fisica, Università di Genova, Via
Dodecaneso 33, I-16146 Genova, Italy} \corauth[cor1]{Tel.:+39 0916234208;
fax: +39 0916162461; e-mail: livigni@fisica.unipa.it}

\begin{abstract}
The magnetic-field-induced variations of the microwave surface resistance have been investigated
in a heavily neutron-irradiated MgB$_2$ sample, in which the irradiation has caused the merging
of the two gaps into a single value. The experimental results have been analyzed in the framework
of the Coffey and Clem model. By fitting the experimental data, we have determined the field dependence
of the depinning frequency, $\omega_0$, at different values of the temperature. Although the pinning
is not particularly effective, the value of $\omega_0$ obtained at low temperatures is considerably
higher than that observed in conventional low-temperature superconductors.
\end{abstract}

\begin{keyword}
Depinning frequency \sep \mgb\ \sep Microwave surface resistance

\PACS 74.25.Ha \sep 74.25.Nf \sep 74.60.Ge

\end{keyword}

\end{frontmatter}

\section{Introduction}
Investigation of fluxon dynamics in type-II superconductors is of great interest for both fundamental and applicative aspects. From the basic point of view, it gives information on the relative magnitude of elastic and viscous forces, which rule the motion regime of the fluxon lattice \cite{gittle,golo,GOLOSOVSKY,TALVA,dulcicvecchio,noiBKBO}. From the technological point of view, it allows determining the magnetic-field-induced energy losses, which have important implication in a large variety of superconductor-based devices \cite{libromw}.

A suitable method to investigate the fluxon dynamics consists in determining the magnetic-field-induced variations of the microwave (mw) surface resistance, $R_s$ \cite{golo}. In the absence of static magnetic fields, the variation with the temperature of the condensed-fluid density determines the temperature dependence of $R_s$. On the other hand, the field dependence of $R_s$ in superconductors in the mixed state is determined by the presence of fluxons, which bring along normal fluid in their cores, as well as the fluxon motion
\cite{gittle,golo,GOLOSOVSKY,TALVA,dulcicvecchio,noiBKBO,CC,BRANDT,noiPRB}. Measurements of the high-frequency em response allow to conveniently investigate the vortex dynamics because they probe the
vortex response at very low currents, when vortices undergo reversible oscillations and are less sensitive to flux-creep processes.

In a fluxon lattice driven by mw currents, the regime of vortex motion is ruled by the relative magnitude of the viscous-drag force, due to the presence of the normal cores, and the restoring-pinning force, which hinders the motion of fluxons. A very important parameter of the vortex dynamics is the depinning frequency, $\omega_0$, which separates two regimes of vortex motion. When the frequency of the driving field, $\omega$, is much larger than $\omega_0$, the viscous-drag force dominates the restoring-pinning force; in this case, the vortex resistivity \cite{golo} is real and the motion of fluxons is highly
dissipative. On the contrary, for $\omega\ll\omega_0$ the vortex resistivity is imaginary and the energy losses are strongly reduced. Measurements of the depinning frequency have been performed in both conventional \cite{gittle,GOLOSOVSKY,gilch} and high-$T_c$ superconductors \cite{GOLOSOVSKY,TALVA,noiBKBO,omega0-HTS}. For temperatures lower enough than $T_c$ and applied magnetic fields smaller enough than $H_{c2}$, conventional superconductors exhibit depinning frequency of the order of MHz, while much higher values ($\gtrsim 10$~GHz) have been reported for cuprate high-$T_c$ superconductors.

Since the first studies on \mgb, different authors have highlighted several anomalies in the field-induced variations of the mw surface impedance, especially at low temperatures and magnetic fields much lower than the upper critical field \cite{shibata,dulcic,nova,isteresiMgB2,Sarti,noiPRBirr}. These studies have established that the standard theories are inadequate to describe the fluxon dynamics in the two-gap \mgb, in wide ranges of temperatures and magnetic fields. This is most likely due to the peculiar properties of the mixed state of \mgb, related to the presence of the two distinct gaps~\cite{eskil,nakai,koshelev,golubovHc2,gure}.

Recently, polycrystalline $\mathrm{Mg}^{11} \mathrm{B}_2$ samples irradiated up to very high neutron fluence have extensively been investigated~\cite{pallecchi,tarantini1,puttiSUST2008,putti2,gonnelli2}. It has been shown that irradiation up to exposure levels of $2\times 10^{18}~\mathrm{cm}^{-2}$ leads to an improvement in the upper critical field and in the field dependence of the critical current density. On further increasing the neutron fluence, all the superconducting properties, such as $T_c$, $H_{c2}$, $J_c$, are reduced. Furthermore, measurements of specific heat, as well as point-contact spectroscopy, have shown that in the sample irradiated at the highest fluence ($1.4 \times 10^{20}~\mathrm{cm}^{-2}$) the two gaps merge into a single value~\cite{putti2,gonnelli2}. Very recently~\cite{puttiSUST2008}, transmission-electron-microscopy studies have shown that neutron irradiation creates nanometric amorphous regions within the \mgb\ crystallites, whose density increases on increasing the neutron fluence. In samples irradiated with neutron fluence $\leq 10^{19}~\mathrm{cm}^{-2}$, such defects act as additional pinning centers. The field dependence of the critical current density observed in these samples has been quantitatively justified by considering the contribution of two pinning mechanisms, one arising from grain boundaries, which is also present in the pristine sample, and the other arising from the defects induced by irradiation. On the contrary, the results of $J_c(B)$ obtained in the samples irradiated with neutron fluence larger than $10^{19}~\mathrm{cm}^{-2}$ have not fully been justified; in this case, the measured $J_c$ values are even lower than those expected from the grain-boundary contribution. A thorough understanding of the pinning mechanisms that come into play in the heavily irradiated samples is not yet achieved.

In this paper, we report a detailed investigation of the magnetic-field-induced variations of the microwave surface resistance of a \mgb\ sample irradiated at the neutron fluence of $1.4\times 10^{20}~\mathrm{cm^{-2}}$. Preliminary results obtained in this sample have shown that the mw losses can be justified in the framework of standard models for vortex dynamics~\cite{noiPRBirr}. Here, we report the results obtained in a wide range of temperatures ($4.2~\mathrm{K}\div T_c$), from which we determine the temperature and the magnetic-field dependencies of the depinning frequency.  The investigation allowed us to determine also the field dependence of the pinning coefficient and the radius of action of the pinning potential.

\section{Experimental apparatus and sample}\label{sec:samples}
The field-induced variations of the mw surface resistance has been investigated in a bulk sample of $\mathrm{Mg}^{11} \mathrm{B}_2$ irradiated at very high neutron fluence ($1.4 \times 10^{20}$ cm$^{-2}$). The procedure for the preparation and irradiation of the sample is reported in detail elsewhere~\cite{tarantini1,putti2}. The sample has been prepared by direct synthesis from Mg (99.999\% purity) and crystalline isotopically enriched $^{11}$B (99.95\% purity), with a residual $^{10}$B concentration lower than 0.5\%. The use of isotopically enriched $^{11}$B makes the penetration depth of the thermal neutrons greater than the sample thickness; this guarantees that the irradiation effects are almost homogeneous over the sample. Several superconducting properties of the sample have been reported in Refs.~\cite{pallecchi,tarantini1,puttiSUST2008,putti2,gonnelli2}. For simplicity and easy of comparison, we label the sample as P6, according to Refs.~\cite{noiPRBirr,pallecchi,tarantini1,puttiSUST2008,putti2,gonnelli2}.   Point-contact-spectroscopy~\cite{gonnelli2} and specific-heat~\cite{putti2} measurements have shown that the neutron-irradiation process determined in this sample a merging of the two gaps into a single value.

The sample has a nearly parallelepiped shape with $w\approx 1.1$ mm, $t\approx 0.8$~mm and $h\approx 1.4$~mm; the effect of the neutron irradiation on this sample led to a worsening of several properties. The superconducting transition is characterized by $T_c^{onset} \approx 9.1$~K and $\Delta T_c\approx 0.3$~K (from 10\% to 90\% of the normal-state resistivity); the residual resistivity ratio is RRR~=~1.1; the critical current density at zero magnetic field is $J_{c0}\approx 3 \times 10^4$~A/cm$^2$, it exhibits a monotonic decrease with the magnetic field, following roughly an exponential law. The upper critical field is isotropic and its value at $T=5$~K is $\mu_0H_{c2}\approx 2~\mathrm{T}$. The measured value of the normal-state resistivity is $\rho_n(40~\mathrm{K})= 130~\mu\mathrm{\Omega~cm}$; however, as suggested by
Rowell~\cite{Rowell}, the real value of the residual normal-state resistivity can be different because of reduction of the effective current-carrying cross-sectional area of the sample due to the grain boundaries. The rescaled value of the residual normal-state resistivity, corrected by the Rowell's criterion, is $\rho_n(T_c)\approx 90~\mu\mathrm{\Omega~cm}$~\cite{tarantini1}.

Although the superconducting transition of the sample is sharp, it results $\Delta T_c/T_c\approx 0.03$. Since the distribution of $T_c$ may affect the temperature dependence of the mw surface resistance near $T_c$, we have determined the $T_c$ distribution function by measurements of the AC susceptibility at 100 kHz. We have found that the first derivative of the real part of the AC susceptibility can be described by a Gaussian distribution function of $T_c$, centered at $T_{c0}=8.5\pm 0.1$~K with $\sigma_{T_c}=0.2\pm 0.05$~K. In the following, we will use this distribution function to quantitatively discuss the results.

\begin{figure}[t]
\centering
\includegraphics[width=7.5cm]{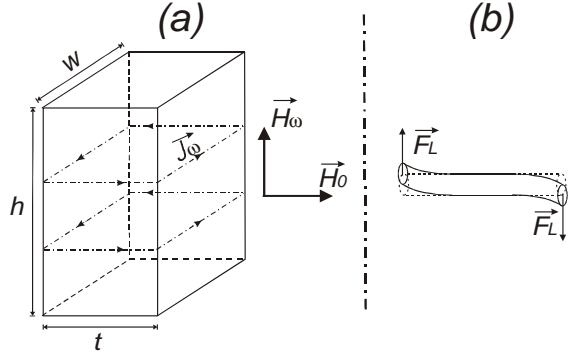}
\caption{(a) Field and current geometry at the sample surface. (b)
Schematic representation of the motion of a flux line.}
\label{sample}
\end{figure}

The mw surface resistance has been measured using the cavity-perturbation technique~\cite{golo}. A copper cavity, of cylindrical shape with golden-plated walls, is tuned in the $\mathrm{TE}_{011}$ mode resonating at $\omega/2\pi \approx 9.6$~GHz. The sample is located in the center of the cavity by a sapphire rod, in the region in which the mw magnetic field is maximum. The cavity is placed between the poles of an
electromagnet which generates DC magnetic fields up to $\mu_0H_0\approx 1$~T. The sample and the field geometries are shown in Fig.~1a; the DC magnetic field, $\emph{\textbf{H}}_{0}$, is perpendicular to the mw magnetic field, $\emph{\textbf{H}}_{\omega}$. When the sample is in the mixed state, the induced mw current causes a tilt motion of the vortex lattice~\cite{BRANDT}; Fig.~\ref{sample}b schematically shows the motion of a flux line.

The surface resistance of the sample is given by\\
\begin{equation*}
    R_s= \Gamma~\left(\frac{1}{Q_L} - \frac{1}{Q_U}\right)\,,
\end{equation*}
where $Q_L$ is the quality factor of the cavity loaded with the sample, $Q_U$ that of the empty cavity and $\Gamma$ the geometry factor of the sample.\\
The quality factor of the cavity has been measured by an hp-8719D Network Analyzer. The surface resistance has been measured as a function of the DC magnetic field, at fixed temperatures. All the measurements have been performed at very low input power; the estimated amplitude of the mw magnetic field in the region in which the sample is located is of the order of $0.1~\mu$T.

\section{Experimental results}\label{experimental}
\begin{figure}[t]
\centering \includegraphics[width=7.5 cm]{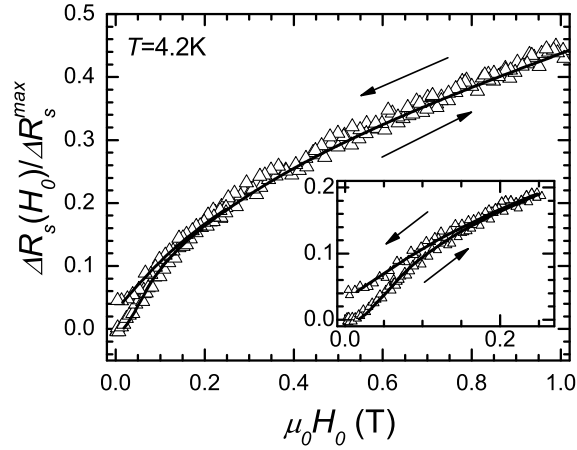}
\caption{Field-induced variations of $R_s$ at $T=4.2$~K. $\Delta R_s(H_0)\equiv R_s(H_0,T)-R_{res}$, where $R_{res}$ is the residual mw surface resistance at $T=2.5$~K and $H_{0}=0$; $\Delta R_s^{max}\equiv R_{n}-R_{res}$, where $R_n$ is the normal-state value of $R_s$ at $T=T_c$. The lines are the best-fit curves obtained, as explained in Ref.~\cite{noiPRBirr}, with $\mu_0 H_{c2}= 1.71$~T, $\omega_0/\omega = 0.67$ and the field dependence of the critical current density reported in Ref.~\cite{tarantini1}. The inset shows a minor hysteresis loop obtained by sweeping $H_0$ from 0 to 0.25~T and back, along with
the best-fit curve.} \label{fig:Rs(H)T=4.2K}
\end{figure}

\begin{figure}[t]
\centering \includegraphics[width=7.5 cm]{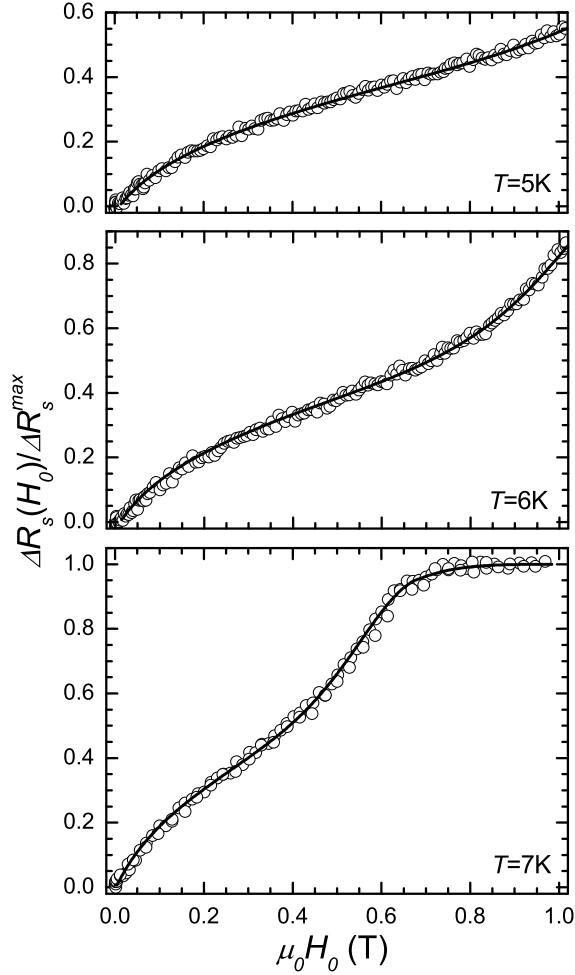}
\caption{Field-induced variations of $R_s$ for the \mgb\ sample, at different values of the temperature. $\Delta R_s(H_0)\equiv R_s(H_0,T)-R_{res}$; $\Delta R_s^{max}\equiv R_{n}-R_{res}$. The lines are the best-fit curves obtained as explained in Sec.~\ref{discussion}.} \label{Rs(H)1}
\end{figure}

\begin{figure}[t]
\centering \includegraphics[width=7.5 cm]{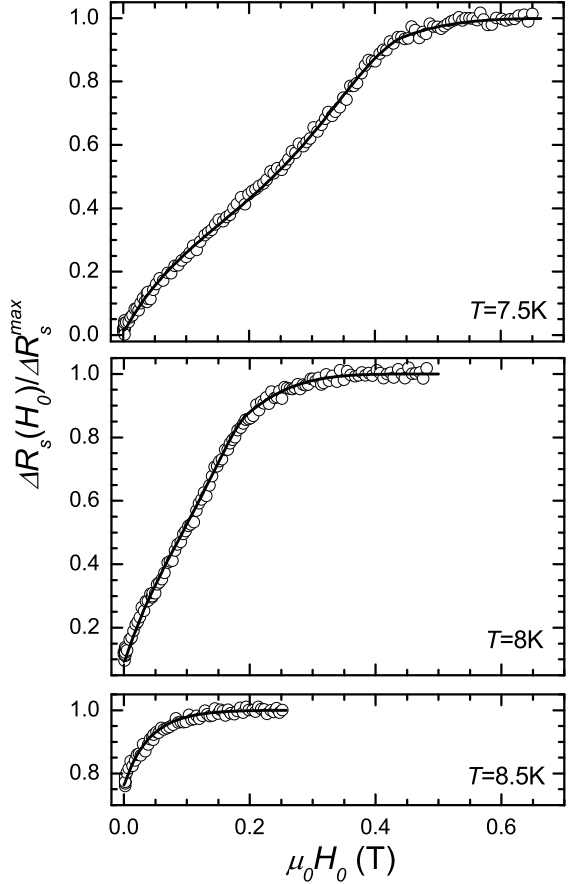}
\caption{Field-induced variations of $R_s$ for the \mgb\ sample, at different values of the temperature. $\Delta R_s(H_0)\equiv R_s(H_0,T)-R_{res}$; $\Delta R_s^{max}\equiv R_{n}-R_{res}$. The lines are the best-fit curves obtained as explained in Sec.~\ref{discussion}.} \label{Rs(H)2}
\end{figure}

The field-induced variations of $R_s$ have been investigated at different temperatures. For each measurement, the sample was ZFC down to the desired value of temperature; the DC magnetic field was increased up to a certain value and, successively, decreased down to zero. Figs.~\ref{fig:Rs(H)T=4.2K},
\ref{Rs(H)1} and \ref{Rs(H)2} show the field-induced variations of $R_s$, at different temperatures. In all the figures, $\Delta R_s(H_0)\equiv R_s(H_0,T)-R_{res}$, where $R_{res}$ is the residual mw surface resistance at $T=2.5$~K and $H_{0}=0$; moreover, the data are normalized to the maximum variation,
$\Delta R_s^{max}\equiv R_{n}-R_{res}$, where $R_n$ is the normal-state value of the surface resistance at $T=T_c$. In the figures, the continuous lines are the best-fit curves obtained by the model described in Sec.~\ref{discussion}.

From Fig.~\ref{fig:Rs(H)T=4.2K} one can see that at $T=4.2$~K the $R_s(H_0)$ curve exhibits a magnetic hysteresis for $H_0$ lower than $\approx 0.18$~T. The hysteresis is ascribable to the different magnetic induction at increasing and decreasing DC fields, due to the critical state of the vortex lattice~\cite{Ji,sridhar,noiisteresi}. These results have been reported and discussed in Ref.~\cite{noiPRBirr}. In order to fit the data, we have determined the $B$ profile inside the sample due to the critical state and calculated a proper averaged value of $R_s(B)$ over the whole sample. The lines in the figure are the best-fit curves.

For $T \geq 5$~K the $R_s(H_0)$ curves are reversible, indicating that at these temperatures the critical-state effects of the fluxon lattice are negligible. However, we would like to remark that, for samples of millimetric size, the sensitivity of our experimental apparatus allows detecting hysteresis in the $R_s(H_0)$ curves for $J_c\gtrsim 10^4~ \mathrm{A/cm}^2$.

\section{Discussion} \label{discussion}

Microwave losses induced by static magnetic fields have been investigated by several
authors~\cite{gittle,golo,GOLOSOVSKY,TALVA,dulcicvecchio,noiBKBO,libromw,CC,BRANDT,noiisteresi,noistatocritico}.
At low temperatures and for applied magnetic fields lower enough than the upper critical field, the main contribution arises from the fluxon motion; however, it has been pointed out that a noticeable contribution can arise from the presence of normal fluid, especially at temperatures near $T_c$ and for magnetic fields of the same order of $H_{c2}(T)$.

In the London local limit, the surface resistance is proportional to the imaginary part of the complex penetration depth, $\widetilde{\lambda}$, of the em field:
\begin{equation}\label{Rs}
    R_s=-\mu_{0}\omega ~\mathrm{Im}[{\widetilde{\lambda}(\omega,B,T)}].
\end{equation}
The complex penetration depth has been calculated in different approximations~\cite{CC,BRANDT}. Coffey and Clem (CC) have elaborated a comprehensive theory for the electromagnetic response of superconductors in the mixed state, in the framework of the two-fluid model of superconductivity~\cite{CC}. The CC theory has
been developed under the assumption that the induction field, $B$, is uniform in the sample; so, it is valid for $H_0 > 2H_{c1}$ whenever the fluxon distribution can be considered uniform within the AC penetration depth.

In the linear approximation, $H_{\omega} \ll\ H_0$, $\widetilde{\lambda}(\omega,B,T)$ expected from the CC model is given by
\begin{equation}\label{lambdat}
    \widetilde{\lambda}(\omega,B,T)=\sqrt{\frac{\lambda^{2}(B,T)+
    (i/2)\widetilde{\delta}_{v}^{2}(\omega,B,T)}
    {1-2i\lambda^{2}(B,T)/\delta _{nf}^{2}(\omega,B,T)}}\, ,
\end{equation}
with
\begin{equation}\label{lamda0}
\lambda(B,T) = \frac{\lambda_0}{\sqrt{[1-w_0(T)][1- B
/B_{c2}(T)]}}\,,
\end{equation}
\begin{equation}\label{delta0}
\delta_{nf}(\omega,B,T) =
\frac{\delta_0(\omega)}{\sqrt{1-[1-w_0(T)][1- B /B_{c2}(T)]}}\,,
\end{equation}
where $\lambda_0$ is the London penetration depth at  $T = 0$, $\delta_0$ is the normal-fluid skin depth at $T = T_c$, $w_0(T)$ is the fraction of normal electrons at $H_0 = 0$; in the Gorter and Casimir two-fluid model $w_0(T)=(T/T_c)^4$.\\
$\widetilde{\delta} _{v}$ is the effective complex skin depth arising from the vortex motion; it depends on the relative magnitude of the viscous and restoring-pinning forces. $\widetilde{\delta} _{v}$ can be written in terms of two characteristic lengths, $\delta_f$ and $\lambda_c$, arising from the contributions of the viscous and the restoring-pinning forces, respectively:
\begin{equation}\label{delta-v}
\frac{1}{\widetilde{\delta}_{v}^{2}}=\frac{i}{2\lambda_{c}^{2}}+\frac{1}{\delta_{f}^{2}}\,,
\end{equation}
where
\begin{equation}\label{lambda-c}
\lambda_{c}^{2}=\frac{B\phi_0}{\mu_0 k_p}\,,
\end{equation}
\begin{equation}\label{delta-f}
\delta_{f}^{2}=\frac{2B\phi_0}{\mu_0 \omega \eta}\,,
\end{equation}
with $k_p$ the restoring-force coefficient, $\eta$ the viscous-drag coefficient and $\phi_0$ the quantum of flux.\\
The effectiveness of the two terms in Eq.~(\ref{delta-v}) depends on the ratio $\omega_0 = k_p/\eta$, which defines the depinning frequency~\cite{gittle}. In terms of $\omega_0$, Eq.~(\ref{delta-v}) becomes
\begin{equation}\label{delta-v(omega)}
\frac{1}{\widetilde{\delta}_{v}^{2}}=\frac{1}{\delta_{f}^{2}}\left(1+
i~\frac{\omega_0}{\omega}\right)\,.
\end{equation}
When $\omega \ll \omega_0$, the fluxon motion is ruled by the restoring-pinning force. On the contrary, for $\omega \gg \omega_0$, the fluxon motion takes place around the minimum of the pinning-potential well and, consequently, the restoring-pinning force is nearly ineffective; so, the contribution of the viscous-drag force predominates and the induced em current makes fluxons move in the flux-flow regime. In the latter case, enhanced field-induced energy losses are expected.

The theory above discussed is strictly valid when $B$ is uniform inside the sample. When fluxons are in the critical state, the assumption of uniform $B$ is no longer valid and the CC theory does not correctly describe the field-induced variations of $R_s$. Recently, we have investigated the field-induced variations of the mw surface resistance in superconductors in the critical state and have accounted for the magnetic hysteresis in the $R_s(H_0)$ curves~\cite{noiisteresi,noistatocritico}. The details of the procedure we have followed to account for the experimental results of Fig.~\ref{fig:Rs(H)T=4.2K}, where the critical-state effects are important, are reported in Refs.~\cite{noiPRBirr}. Since $R_s(H_0)$ in the investigated sample does not show hysteresis in a wide range of temperatures, we do not discuss here on this procedure.

The expected value of the normalized surface resistance depends on several parameters, such as the
ratio $\lambda_0/\delta_0$, the temperature dependence of the normal-fluid density $w_0(T)$, $H_{c2}(T)$, the depinning frequency and its field dependence. However, $\lambda_0/\delta_0$ and the temperature dependence of the normal-fluid density determine the value of $R_s(T)$ at $H_0=0$. In Ref.~\cite{noiPRBirr} we have shown that the $R_s(T)$ curve at $H_0=0$ can be quite well justified assuming valid the Gorter and Casimir two-fluid model, with $\lambda_0/\delta_0$ values ranging from 0.04 to 0.15, provided that distribution of $T_c$ in the sample is taken into account. The large uncertainty of $\lambda_0 / \delta_0$ is due to the fact that the $T_c$ distribution broadens the $R_s(T)$ curve, hiding the $\lambda_0 / \delta_0$ effects.

At fixed temperature, the expected field-induced variations of $R_s$ depend on $\omega_0(B)$ and $H_{c2}(T)$. The temperature dependence of $H_{c2}$ has been reported in Refs.~\cite{noiPRBirr,tarantini1}; it can be described by the law
\begin{equation}\label{Hc2(T)}
H_{c2}(T)=H_{c20}[1-(T/T_c)^{\alpha}]\,,
\end{equation}
with $\mu_0 H_{c20}=(2.2\pm0.2)$~T, $\alpha=1.9\pm0.3$ and
$T_c=(8.9\pm0.2)$~K.\\
By using this relation for $H_{c2}(T)$, it is possible to determine the field dependence of the depinning frequency by fitting the experimental isothermal $R_s(H_0)$ curves.

\begin{figure}[b]
\centering \includegraphics[width=7.5 cm]{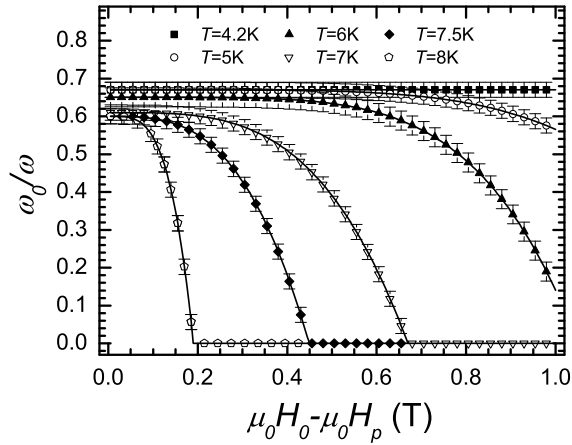}
\caption{Magnetic field dependence of the depinning frequency, obtained by fitting the experimental $R_s(H_0)$ curves, at different temperatures.} \label{Omega0}
\end{figure}

For $T\geq 5$~K, the $R_s(H_0)$ curves do not show hysteresis, suggesting that the effects of the critical state are negligible. Since on increasing the temperature the effects of the distribution of $T_c$ become more and more important, to fit the experimental data we have averaged the expected $R_s(H_0)$ curves [calculated by Eqs.~(\ref{Rs}-\ref{delta0})] over the $T_c$ distribution; have used Eq.~(\ref{Hc2(T)}) for $H_{c2}(T)$; have taken $\omega_0$ as parameter dependent on $H_0$. Moreover, we have used the following approximate expression for the magnetization:
\begin{equation}\label{m}
    M=-H_p+ \frac{H_p}{H_{c2}-H_p}(H_0-H_p)\,
\end{equation}
and consequently
\begin{equation}\label{B}
    B=\mu_0 \left(1+\frac{H_p}{H_{c2}-H_p}\right)(H_0-H_p)\,,
\end{equation}
where $H_p$ is the first-penetration field.\\
$H_p$ can be directly deduced from the experimental curves, measuring the applied magnetic field at which $R_s$ starts to increase; its temperature dependence has been reported in Ref.~\cite{noiPRBirr}. The best-fit curves of $R_s(H_0)$ are reported in Figs.~\ref{fig:Rs(H)T=4.2K}, \ref{Rs(H)1} and \ref{Rs(H)2}; the field dependence of $\omega_0/\omega$, by which the best fit has been obtained, is reported in Fig.~\ref{Omega0}, at different temperatures. We remark that we have investigated the field-induced variations of $R_s$ down to $T=2.5$~K; the results of the depinning frequency for $T<4.2$~K are not reported here because, within the experimental uncertainty, they are the same of those obtained at $T=4.2$~K.

Considering the frequency of the mw field ($\omega/2\pi \simeq 9.6$~GHz), from $\omega_0/\omega$ we obtain that the depinning frequency at low fields is $\omega_0/2\pi \approx 6$~GHz. For $T\leq 5$~K, $\omega_0$ is independent of $B$ in a wide field range; on increasing the temperature, this field range shrinks. Moreover, for applied fields larger than a threshold value, dependent on $T$, the fluxon lattice moves in the flux-flow regime ($\omega_0 \ll \omega$); on increasing the temperature, this threshold field decreases; eventually, at $T>8$~K the fluxon lattice moves in the flux-flow regime in the whole field range investigated.

Two regimes of vortex pinning can be identified: individual pinning and collective pinning. Individual pinning is realized at low fields, when there are few vortices and many pinning sites per vortex. In this regime, $\omega_0$ is expected to be independent of $B$. Collective pinning is realized at higher fields when the vortex concentration is high and there are many vortices per pinning site; in this regime, $\omega_0$ gets lower values and depends on $B$~\cite{golo}. Our experimental results indicate that
for $T \lesssim T_c / 2$ individual pinning is realized in almost the whole field range investigated.

\begin{figure}[t]
\centering \includegraphics[width=7.5 cm]{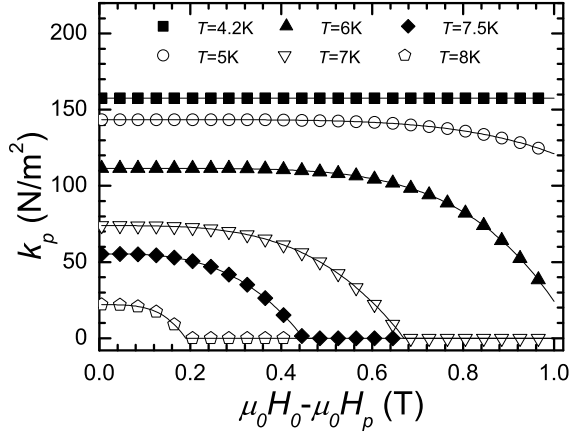}
\caption{Deduced field dependence of the pinning coefficient $k_p$, at different temperatures.} \label{Kp(B)}
\end{figure}

Since the field-induced mw losses depend on the relative magnitude of the viscous and restoring-pinning forces, our analysis does not allow to obtain independently the coefficients $k_p$ and $\eta$, but only the ratio $\omega_0 = k_p/\eta$. However, the investigated sample has shown properties that can be quite well accounted for by conventional models; so, one can deduce $k_p$ by supposing valid the Bardeen-Stephen
relation~\cite{BS}
\begin{equation}\label{Bard-Step}
    \eta(T)=\frac{\phi_0 \mu_0H_{c2}(T)}{\rho_n}\,,
\end{equation}
where $\rho_n$ is the normal-state resistivity at $T=T_c$.\\
Both $\rho_n(T_c)$ and $H_{c2}(T)$ of the investigated sample have been already determined: $\rho_n(T_c)\approx 90~\mu\mathrm{\Omega~cm}$~\cite{tarantini1} and $H_{c2}(T)$ is given by Eq.~(\ref{Hc2(T)}). The deduced field dependence of $k_p$ is shown in Fig.~\ref{Kp(B)}, at different values of the temperature.

An upper limit of the pinning constant, $k_p^{max}$, can be obtained by equating the energy density per unit length of the vortex core, $B_c^2 \xi^2 \pi / 2\mu_0$, to the elastic stored energy density per unit length of the vortex core, $k_p^{max} \xi ^2 / 2$~\cite{Wu}. It follows:
\begin{equation}\label{KpMax}
    k_p^{max}=\frac{\pi B_c^2}{\mu_0}\,,
\end{equation}
where $B_c$ is the thermodynamic critical field.\\
At low temperatures, from $B_{c1}$ and $B_{c2}$ we estimate that the thermodynamic critical field is of the order of 100~mT; so,
from Eq.~(\ref{KpMax}) we estimate $k_p^{max}$ to be of the order of $10^4$~N/m$^2$. The values of $k_p$ we have obtained from the experimental data are two orders of magnitude lower than $k_p^{max}$ deduced from Eq.~(\ref{KpMax}); so, we infer that pinning is not particularly effective in the investigated sample. This finding is consistent with the low $J_c$ value responsible for the weak hysteretic behavior of $R_s$ reported in Fig.~\ref{fig:Rs(H)T=4.2K}.

From simultaneous measurements of the pinning constant and critical current density it is possible to estimate the average radius of action, $r_p$, of the pinning potential, $U(r)$. The pinning constant, $k_p=d^2U/dr^2$, is related to the critical current density; since $J_c=\phi_0^{-1}dU/dr|_{r_p}$, one obtains $J_c=k_pr_p/ \phi_0$~\cite{golo}. By using the value of $J_c(5~\mathrm{K})$ reported in Ref.~\cite{tarantini1} and the value we obtained for $k_p$, it results $r_p\approx 4$~nm. At applied fields of $\approx 1$~T, the estimated distance between vortices is $\sqrt{\phi_0/B} \approx 45$~nm; so, the deduced value of $r_p(5~\mathrm{K})$ is much smaller than the distance between vortices. This finding confirms that, in our sample, for $T \leq5$~K individual pinning is realized in almost the whole field range investigated, consistently with the field-independent depinning frequency obtained at these  temperatures (see Fig.~\ref{Omega0}).

It is widely accepted that in pristine MgB$_2$ bulk samples the pinning mechanism is ruled by grain boundaries \cite{pallecchi,tarantini1,puttiSUST2008,eisterer,SUST-Perkins}.
Recently, it has been shown that neutron irradiation introduces defects in the form of amorphous regions of mean diameter $\sim 4$~nm~\cite{puttiSUST2008}, uniformly distributed within the crystallites; the defect density increases on increasing the neutron fluence. Studies on the field dependence of the critical current density~\cite{pallecchi,puttiSUST2008} have suggested that at moderate neutron-fluence levels ($\leq 1 \times 10^{19}$~cm$^{-2}$) these defects give a further contribution to the pinning, leading
to an improvement of the critical current density, with respect to the values expected from grain-boundary pinning. The same effect has not been observed in samples irradiated with higher fluences; in this case, the measured $J_c$ values are even lower than those expected by properly rescaling the grain-boundary contribution. Most likely, this is due to the different coherence length of the different samples: in samples exposed to neutron fluence $\leq 1 \times 10^{19}$~cm$^{-2}$ the defect dimension matches with the coherence length; in samples exposed to higher fluences the coherence length is larger than the defect size and, consequently, the defects are not effective for the fluxon pinning. The results we have obtained in the heavily irradiated sample confirm this conclusion; indeed, we have obtained small values of the pinning coefficient. So, despite the high concentration of defects, they do not contribute positively to the pinning.

The depinning frequency in the investigated sample is considerably higher than the values reported in the literature for conventional superconductors; for example, the depinning frequency in bulk niobium, which has comparable values of $H_{c2}$, is less than $10^8$~Hz~\cite{golo}. Since the values we have obtained for the pinning constant $k_p$ are considerably lower than the upper limit $k_p^{max}$, the high value of $\omega _0$ cannot surely be ascribed to strong pinning effects; so, we infer that it is ascribable to a low value of the viscosity coefficient. On the other hand, in this sample the residual normal-state resistivity is $\rho(T_c)= 90~\mu\Omega$~cm; this high value of $\rho(T_c)$ has been ascribed to a reduced value of the electron mean free path due to the presence of the defects induced by the neutron irradiation~\cite{tarantini1}. We suggest that it is just the high value of the normal-state resistivity responsible for the small viscosity coefficient that, in turn, gives rise to the high value of the depinning frequency.
\section{Conclusion}
We have measured the magnetic-field-induced variations of the mw surface resistance in a heavily neutron-irradiated Mg$^{11}$B$_2$ sample, in which the two gaps merged into a single value. The field dependence of $R_s$, at different values of the temperature, have been discussed in the framework of
the Coffey and Clem model, with the temperature dependence of the normal-fluid density expected from the Gorter and Casimir two-fluid model.

By fitting the experimental data, we have determined the magnetic-field dependence of the depinning frequency at different temperatures. We have found that, for $T\lesssim T_c/2$ the depinning frequency $\omega_0$ does not depend on the magnetic field, indicating that individual pinning is realized in the whole field range investigated; on increasing the temperature, the range of $H_0$ in which individual pinning occurs shrinks and $\omega_0$ is field dependent above a certain threshold value of the applied field, depending on $T$. By supposing valid the Bardeen-Stephen relation for the viscosity coefficient, we have deduced the field dependence of the pinning coefficient at different temperatures; moreover, considering the value of the critical current density, we have deduced the radius of action of the pinning potential. Although the neutron-irradiation process created a high density of defects, our results show that the pinning is not particularly effective, consistently with the relatively low value of the critical current density reported for this sample; this finding is most likely due to the fact that the coherence length is larger than the mean size of the defects. Nevertheless, the deduced value of the depinning frequency is considerably higher than that reported for conventional SC, as bulk Nb. We suggest that this high value of $\omega_0$ is due to the high value of the normal-state resistivity in the investigated sample, which is due to the reduced value of the electron mean free path because of the presence of the defects.


\section*{Acknowledgements}
The authors are very glad to thank C. Ferdeghini and M. Putti for their interest to this work and helpfull suggestions; G. Lapis and G. Napoli for technical assistance.


\end{document}